\documentclass[journal]{IEEEtran}

\usepackage[latin1]{inputenc}
\usepackage{times,amsmath}
\usepackage{amssymb}
\usepackage{pstool}
\usepackage{subfigure}
\usepackage{multirow}
\usepackage{enumerate}
\usepackage{graphicx}
\usepackage{MnSymbol}
\usepackage{stfloats} 
\usepackage[table]{xcolor}
\usepackage[square, comma, sort&compress, numbers]{natbib}
\usepackage{nohyperref}
\usepackage{epstopdf}
\usepackage{comment}
\usepackage[ruled,noend, vlined,linesnumbered]{algorithm2e}
\usepackage{amsthm}

\usepackage{soul}

\usepackage[T1]{fontenc} 

\usepackage{bbm}

\graphicspath{ {Figures/} }

\begin{document}
\title{Expanding Boundaries: Cross-Media Routing for Seamless Underwater and Aerial Communication}

\author{\IEEEauthorblockN{
Waqas Aman\IEEEauthorrefmark{2}, 
Flavio Giorgi\IEEEauthorrefmark{1}, 
Giulio Attenni\IEEEauthorrefmark{1}, 
Saif Al-Kuwari\IEEEauthorrefmark{2}, 
Elmehdi Illi\IEEEauthorrefmark{2},
Marwa Qaraqe\IEEEauthorrefmark{2},\\
Gaia Maselli \IEEEauthorrefmark{1},
Roberto Di Pietro\IEEEauthorrefmark{2}}


}
\maketitle
\vspace{-0.2in}
\begin{abstract}
The colossal evolution of wireless communication technologies over the past few years has driven increased interest in its integration in a variety of less-explored environments, such as the underwater medium. In this magazine paper, we present a comprehensive discussion on a novel concept of routing protocol known as cross-media routing, incorporating the marine and aerial interfaces. In this regard, we discuss the limitation of single-media routing and advocate the need for cross-media routing along with the current status of research development in this direction. To this end, we also propose a novel cross-media routing protocol known as bubble routing for autonomous marine systems where different sets of AUVs, USVs, and airborne nodes are considered for the routing problem. We evaluate the performance of the proposed routing protocol by using the two key performance metrics, i.e., packet delivery ratio (PDR) and end-to-end delay. Moreover, we delve into the challenges encountered in cross-media routing, unveiling exciting opportunities for future research and innovation. As wireless communication expands its horizons to encompass the underwater and aerial domains, understanding and addressing these challenges will pave the way for enhanced cross-media communication and exploration. 
\end{abstract}


\vspace{-0.2cm}
\section{Introduction}
\label{sec:intro}

\begin{figure}
    \centering
    \includegraphics[scale=0.25]{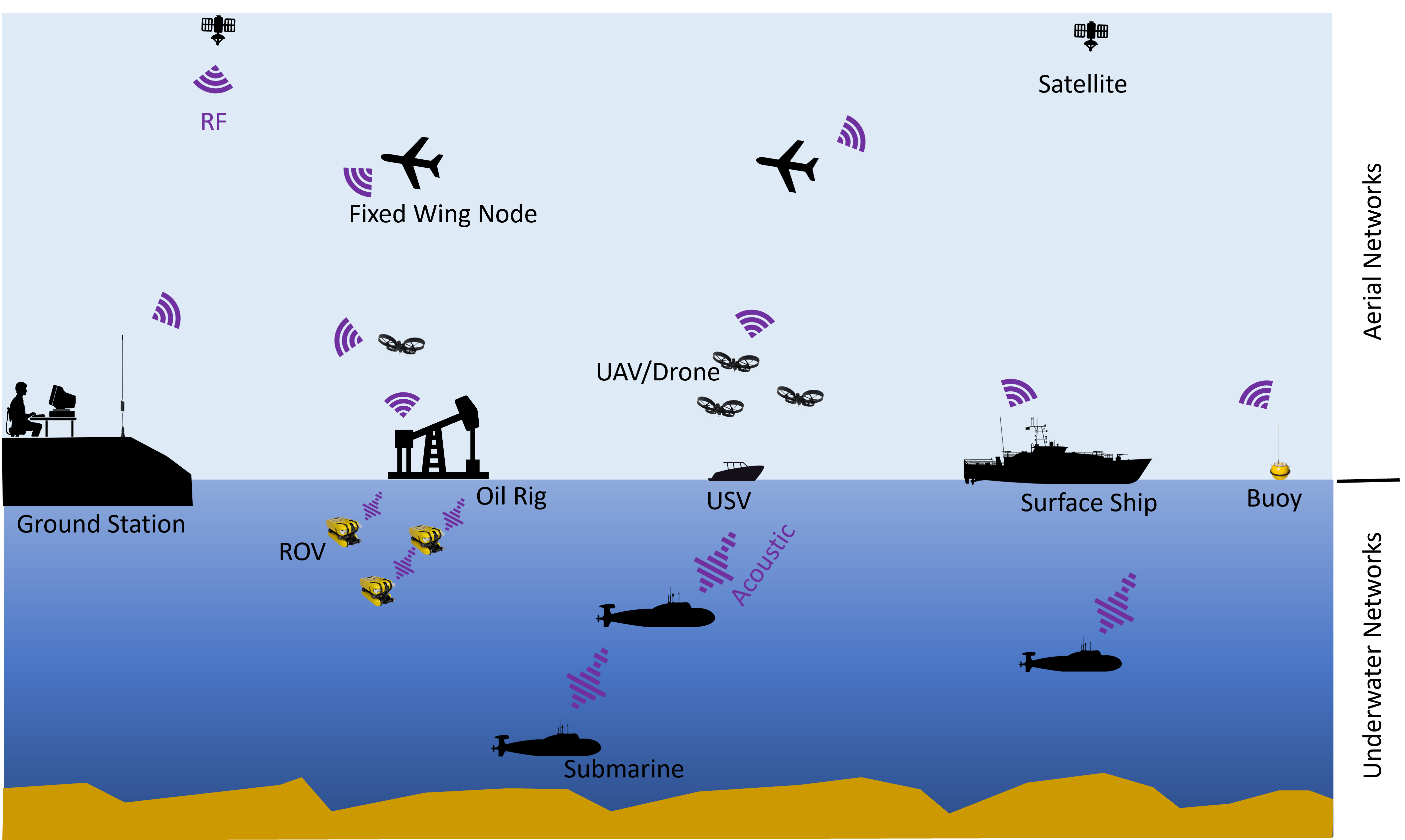}
    \caption{Forging the Future: A Multifaceted Mesh of Subaquatic and Skies - Illustrating the Intricate Networks of AMS.  Typically, submarines and remote operating vehicles (ROVs)/autonomous underwater vehicles (AUVs) are considered as underwater nodes, oil rigs, unmanned surface vehicles (USVs), surface ships, and Bouys as surface nodes, and unmanned aerial vehicles (UAVs)/drones, Fixed wing nodes (also known as airplanes or flying jets) and satellites are considered as airborne nodes. These all are connected directly or via relay links to the ground station.  }
    \label{fig:AMS}
\end{figure}

Autonomous Marine Systems (AMSs) represent a fast-growing research field that combines heterogeneous devices (e.g., Unmanned Aerial/Surface/Underwater Vehicles and Sensor Nodes) and communication technologies (e.g., Radio Frequency, acoustic signals, magnetic induction, Reconfigurable Intelligent Surfaces) to enable the development of self-organizing networks in the marine environment. 
These systems are indeed set to play a crucial role as technology advances due to their potential to increase operational efficiency, cost-effectiveness, and safety in a variety of applications, including oceanic research, environmental monitoring, underwater exploration, surveillance, and offshore industry operations \cite{zolich2019survey}. Fig. \ref{fig:AMS} envisions a typical architecture of AMSs.
During the last few years, there has been a rapid and considerable growth of Underwater Wireless Sensors Networks (UWSNs) and Flying Ad-hoc Networks (FANETs), this expansion brought many innovations, and nowadays, both UWSNs and FANETs are widely used separately in many different scenarios. In particular, UWSN applications require deployment in remote areas where it is extremely tough, if not impossible, to communicate directly with a ground station. As pointed out in Nomikos et al. \cite{surveynomikos}, to address this problem, it is possible to use UAVs to gather the information from the UWSN and offload them toward a ground station exploiting UAVs connectivity to build a radio communication channel. For example, in Ludvigsen et al. \cite{ludvigsencrossmedia}, they proposed a system used to map the sea bed that includes heterogeneous vehicles: one USV, two UAVs, and one UUV. The system can reduce dependency on the support vessel by utilizing the USV as a bridge between the underwater network and the overwater network using both acoustic and radio frequency modems. The UAVs instead can gather this data and send them over a longer distance to the support vessel or to a ground station. Wu et al. \cite{cooperativeWu} focuses on a search-and-track (SAT) mission for an underwater target, combining an unmanned aerial vehicle (UAV), an unmanned surface vehicle (USV), and an unmanned underwater vehicle (UUV) and developing a cooperative strategy and path planning algorithm for the heterogeneous system including constraints on communication and detection mechanism.

By combining aerial and underwater sensing, these networks provide enhanced situational awareness in maritime environments. Indeed, these networks combine the perspectives of the aerial and underwater domains to enable the collection of comprehensive data. Aerial devices are able to gather data over a wide geographical area, monitor surface conditions, and take high-resolution pictures. Surface devices can provide more fine-grained information about surface conditions through in-situ measurements. Underwater devices can carry out activities at different depths. Integrating these tools enables a deeper comprehension of underwater structures, complex ecosystems, and disaster-affected places.
In order to unleash the full potential of collaborative work, ensure dependable data transmission across diverse environments, and obtain improved response capabilities communication protocols play a vital role.

Offering enhanced capabilities for monitoring, detecting, and responding to various activities and events in the marine environment, AMSs are appealing for marine surveillance applications. 
In this type of application, it is reasonable to presume that deployed devices are interested in communicating with a ground station that collects all the gathered data in order to provide a centralized view of the entire environment. 
Typically, in AMS, acoustic communication technology is preferred for underwater due to its long communication range nature, and RF for above the water due to its promising results. Therefore, the focus of this work is on AMS which employs RF for above-the-water nodes and acoustic for underwater nodes.  


\vspace{-0.2cm}

\section{Overview of Cross Media Routing}\label{sec:related_works}
In this section, first, we advocate \textit{why traditional or single media routing is not sufficient} for heterogeneous systems like AMS. Next, we emphasize the need for cross-media routing for AMS. Moreover, we discuss the current development in the field and highlight their limitations.
\vspace{-0.2cm}

\subsection{Limitations of Single Media Routing for AMS}
Single-media routing refers to the practice of using a routing algorithm designed for a specific communication medium in cross-media or single-medium networks. It may perform well within the specific medium, but it is not suitable for cross-media networks due to several reasons, e.g., lack of interoperability, heterogeneous communication characteristics, optimal resource utilization, fault tolerance, and scalability and adaptability. We discuss them one by one below:   

\textbf{Heterogeneous Communication Characteristics:} Different communication media have distinct characteristics such as bandwidth, propagation delays, signal attenuation, and energy constraints. Single media routing algorithms are tailored to optimize performance within a specific medium, but they may not consider the unique challenges and limitations of other media. Consequently, they may not be able to effectively handle the heterogeneity of cross-media networks.

\textbf{Lack of Interoperability:} Cross-media networks require seamless communication and data exchange between different media. Single media routing algorithms are typically not designed to accommodate communication across multiple media or to facilitate smooth transitions between them. They may lack the necessary mechanisms to handle media handoffs, adapt to varying channel conditions, or exploit the advantages of different media.

\textbf{Optimal Resource Utilization:} Cross-media networks offer the potential for utilizing multiple communication channels to optimize resource utilization. For example, underwater acoustic networks may have limited bandwidth and higher propagation delays, while satellite or terrestrial networks can offer higher bandwidth and lower latency. Cross-media routing enables efficient utilization of available resources by dynamically selecting the most appropriate media for data transmission. Single media routing algorithms cannot exploit such opportunities for resource optimization.

\textbf{Resilience and Fault Tolerance:} Cross-media networks often require robustness and fault tolerance to overcome challenges such as link failures, interference, or congestion. Single media routing algorithms may lack the mechanisms to dynamically reroute traffic across different media in response to network failures or changing conditions. Cross-media routing algorithms, on the other hand, can provide alternative paths and ensure reliable communication by leveraging multiple media options.

\textbf{Scalability and Future Adaptability:} As cross-media networks expand in size and complexity, scalability becomes crucial. Single media routing algorithms may not scale well when applied to large-scale cross-media networks. Additionally, future advancements and changes in communication technologies may introduce new media or communication channels. Cross-media routing algorithms are designed to be adaptable and future-proof, allowing for seamless integration of new media into the network.

In the context of AMSs, there are multiple communication channels or media that may be involved, such as underwater acoustic networks, satellite communications, and terrestrial networks. Cross-media routing enables effective communication and data transfer between these various media to facilitate the exchange of information and control signals in an AMS.
\vspace{-0.2cm}

\subsection{Need for Cross-media Routing}
Cross-media routing is essential in AMS for several reasons:

\textbf{Integration of Multiple Communication Channels:} AMSs often rely on various communication channels such as underwater acoustic networks, satellite communications, and terrestrial networks. Cross-media routing enables seamless integration and efficient data transfer between these different channels, allowing the system to leverage the advantages of each medium.

\textbf{Enhancing Communication Reliability:} By employing cross-media routing, AMSs can improve communication reliability. If one communication channel experiences disruptions, such as underwater acoustic networks being affected by environmental conditions or limited bandwidth, cross-media routing allows for dynamic switching to alternate channels, such as satellite or terrestrial networks, to maintain uninterrupted communication.

\textbf{Optimizing Data Transfer:} Different communication channels have varying bandwidths, propagation delays, and reliability. Cross-media routing algorithms consider these factors to optimize data transfer based on the characteristics of each medium. It ensures efficient utilization of available bandwidth, minimizes latency, and maximizes the successful delivery of critical data in AMSs.

\textbf{Extending Communication Range:} Each communication medium has its limitations regarding range and coverage. Cross-media routing enables the extension of the communication range by bridging the gaps between different media. For instance, underwater vehicles equipped with acoustic communication systems can relay data to surface buoys with satellite communication capabilities, which can then transmit the data to remote command centers or other platforms.

\textbf{Enabling Remote Monitoring and Control:} Cross-media routing facilitates remote monitoring and control of AMSs. It allows operators and researchers to access real-time data, send commands, and receive feedback from the system regardless of their location. This capability is particularly valuable for applications such as underwater exploration, environmental monitoring, and offshore operations.

Overall, cross-media routing in AMSs enables efficient and reliable communication, extends the system's reach, and enhances the overall performance and capabilities of the system by integrating multiple communication channels.
\vspace{-0.2cm}

\subsection{Current Status}
We found two partially related recently reported papers on the topic of discussion \cite{crossmedia_ding,verma2022towards}. Specifically, Ding et al. \cite{crossmedia_ding} propose a so-called cross-media routing solution that encompasses a LEACH-based clustering algorithm that reduces the number of nodes participating in forwarding and a Vector-based routing protocol (VBCM) that enables nodes to select low-delay links to communicate which is based on the self-adaptive selection of tasks. 
Next, Verma et al. \cite{verma2022towards} presents an Energy-Efficient UAV-assisted routing mechanism for 5G Internet of Underwater Things (IoUT). They deploy a fixed number of energy harvesting-enabled sensor nodes that act as gateways to collect data from the underwater nodes. A UAV hovers over the target area to gather the information from the gateway nodes to send the collected data to the ground station through the 5G network. To improve the performance, they used a cluster-based routing method for the underwater sensor nodes wherein the cluster head is selected through \textit{Improved-Tunicate Swarm Algorithm} (I-TSA) exploiting  parameters such as residual energy, distance from the energy harvesting enabled nodes, network's average energy, and fault-tolerant characteristic of a node. 

\textit{The above works are superficial and do not encounter the main challenges faced by the system. Specifically, one of the main challenges, i.e., bandwidth and data rate differences at the partition line of two different networks has been totally kept out of consideration. Rest, other challenges which create new directions for future work are not been covered, or discussed, which we do in Section \ref{sec:ORI}.
}

\vspace{-0.2cm}

\section{Proposed Mechanism}
\label{sec:proposed_routing}

\begin{figure}[htb!]
    \centering
    \includegraphics[width=0.3\textwidth]{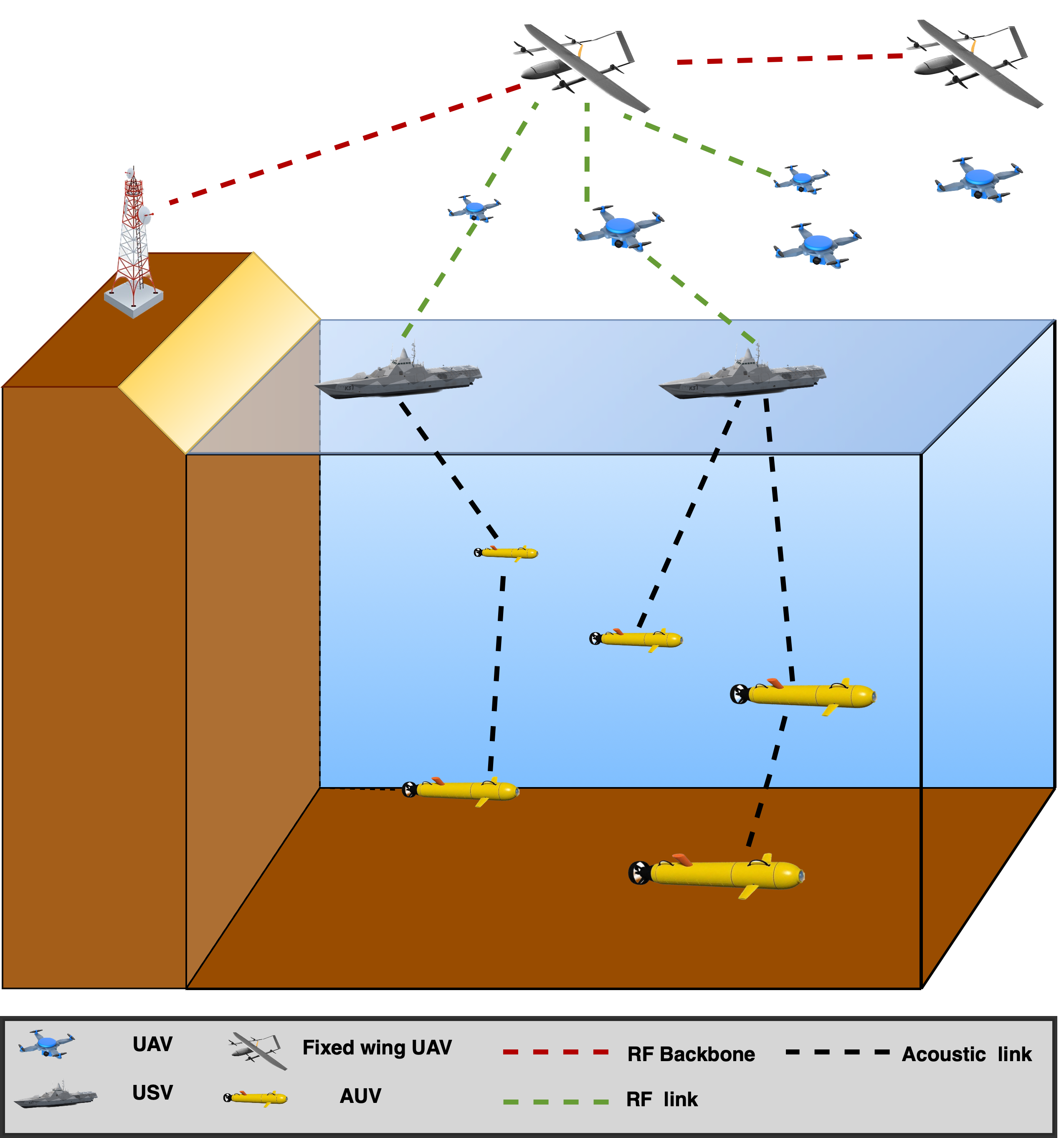}
    \caption{System Model}
    \label{fig:system_model}
\end{figure}
Keeping the need and importance of cross-media routing for AMS, we propose a novel routing mechanism for AMS. In this regard, we consider a system presented in Fig. \ref{fig:system_model} comprises of a set of AUVs, USVs, UAVs, fixed wings nodes, and a ground station. Specifically, we have a three different layer system: the first layer is the \textbf{underwater layer}, which is made up of devices designed to operate underwater (i.e., UUVs, namely Unmanned Underwater Vehicles, and moored nodes). 
The second layer is the \textbf{surface layer}, which is made up of devices designed to operate on the surface of water-bodies (i.e., USV, namely Unmanned Surface Vehicles, SV, namely Surface Vehicles, and moored buoys).
The third and last layer is the \textbf{aerial layer}, which is made up of devices designed to operate in the air (i.e., UAVs, namely Unmanned Aerial Vehicles, Remotely Piloted AVs). We assume that USVs and airborne nodes are facilitating the nodes of interest (AUVs) by presenting themselves as relay nodes. Note that the surface layer is crucial for underwater-to-ground communications because the elements belonging to this layer act as gateways toward the over-water layer. 

We envisage that the communications among nodes happen in two segments: the underwater segment, which enables acoustic signals communications among underwater and surface nodes; and the aerial segment, which enables radio-frequency signals communications among surface, aerial nodes, and the ground station. \textit{The challenge of bandwidth and data rate differences is coped with in a heuristic way, i.e., a time division approach is introduced where for a fraction of the time AUVs transmit their packets using the CSMA/CA approach as medium access control as per the JANUS standard \cite{Janus:Ucomms:2014}, and for the rest of the whole time frame, USVs communicates with the airborne nodes and so on}. The RF network also use CSMA/CA for medium access control.



Next, we present a novel routing mechanism that enables \textit{node-to-sink} communication in the network described above. Firstly, we need to describe the topology of the network, which is dynamic over time. Then, we describe the process through which the nodes maintain and spread information about the local topology introducing a buffer saturation avoidance mechanism. Finally, we describe the process through which each node selects a relay.

\begin{figure}
    \centering
    \includegraphics[width=0.3\textwidth]{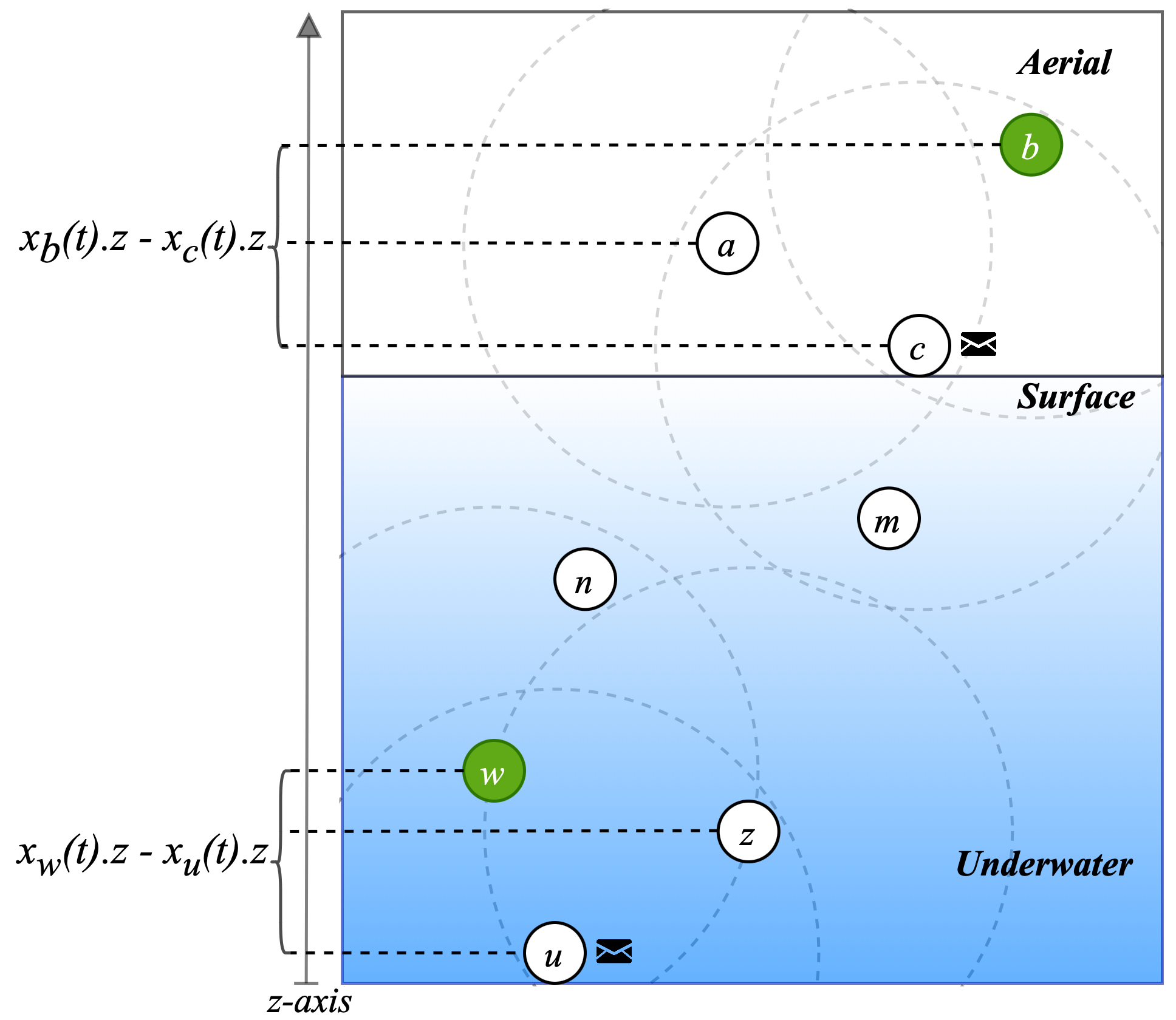}
    \caption{Node $u$ wants to send a packet toward one of its neighbors and selects $w$ (green node) because it provides the greatest z-axis advancement. The node $c$ wants to send a packet and also in this case it chooses the node with the greatest advancement which is $b$}
    \label{fig:bubble}
\end{figure}

\vspace{-0.2cm}

\subsection{Topology}
The network proposed has a dynamic topology. Let $V$ be the set of nodes
(the underwater nodes, the surface nodes, and the aerial nodes). Considering the states of a node over time, we can describe its motion in a 3D space, where $p_{u}(t)$ represents the coordinate of the node at time $t$, $v_{u}(t)$ represents the speed vector at time $t$ (namely, the speed in each of the three components), and $\hat{p}_{u}(t)$ represents coordinates of the next way-point toward the node is moving at time $t$. The state $s_u(t)$ also includes the buffer occupancy of $u$ at time $t$, namely, $b_u(t)$ which is measured as the number of packets contained in the buffer at time $t$. 
This state formulation is applicable to both fixed and mobile nodes. Indeed, if a node is fixed, its speed is 0 and its coordinates never change.

\vspace{-0.2cm}

\subsection{Network setup}
To be able to select a relay, each node must be aware of its neighbors.
To do so, we envisage a mechanism that allows nodes to set up and maintain forwarding tables.
Each node can enter the network broadcasting an \textit{hello packet}. Every node that receives a hello packet uses the information contained in it to build and update its forwarding table. We also assume that a node can also decide to temporarily, or permanently, leave the network broadcasting a \textit{goodbye packet}. Each node should take this decision autonomously considering the expected buffer saturation at a given time.

\subsection{Relay selection}
The selection of a packet next hop is a crucial operation in the routing process.
We propose two relay selection methods: \textit{backbone relay selection}, and \textit{bubble relay selection}.

\paragraph{Backbone relay selection} This mechanism is meant to be implemented in fixed-wing UAVs, which constitute the top layer of the network topology. Once packets have reached these nodes they should be forwarded only to other fixed-wing UAVs or directly to the ground station. Note that this is possible due to the high-power long-range transmitter that this type of UAVs can be equipped with. Therefore, the relay selection is either the top layer node with the greatest advancement towards the ground station or the ground station itself if it is in the communication range.


\paragraph{Bubble Routing relay selection} This mechanism is applied to forward packets up until they are received by the top layer. The goal of this procedure is to move the packets toward the top layer of the network topology. Each node selects the node in its neighborhood with the greatest advancement towards the sky (i.e., the \textit{z-axis}) as it is possible to see in Figure \ref{fig:bubble}. As a tie-breaking rule, the node with the greatest advancement toward the ground station is picked as a relay. 

\begin{figure}
    \centering
    \includegraphics[scale=0.25]{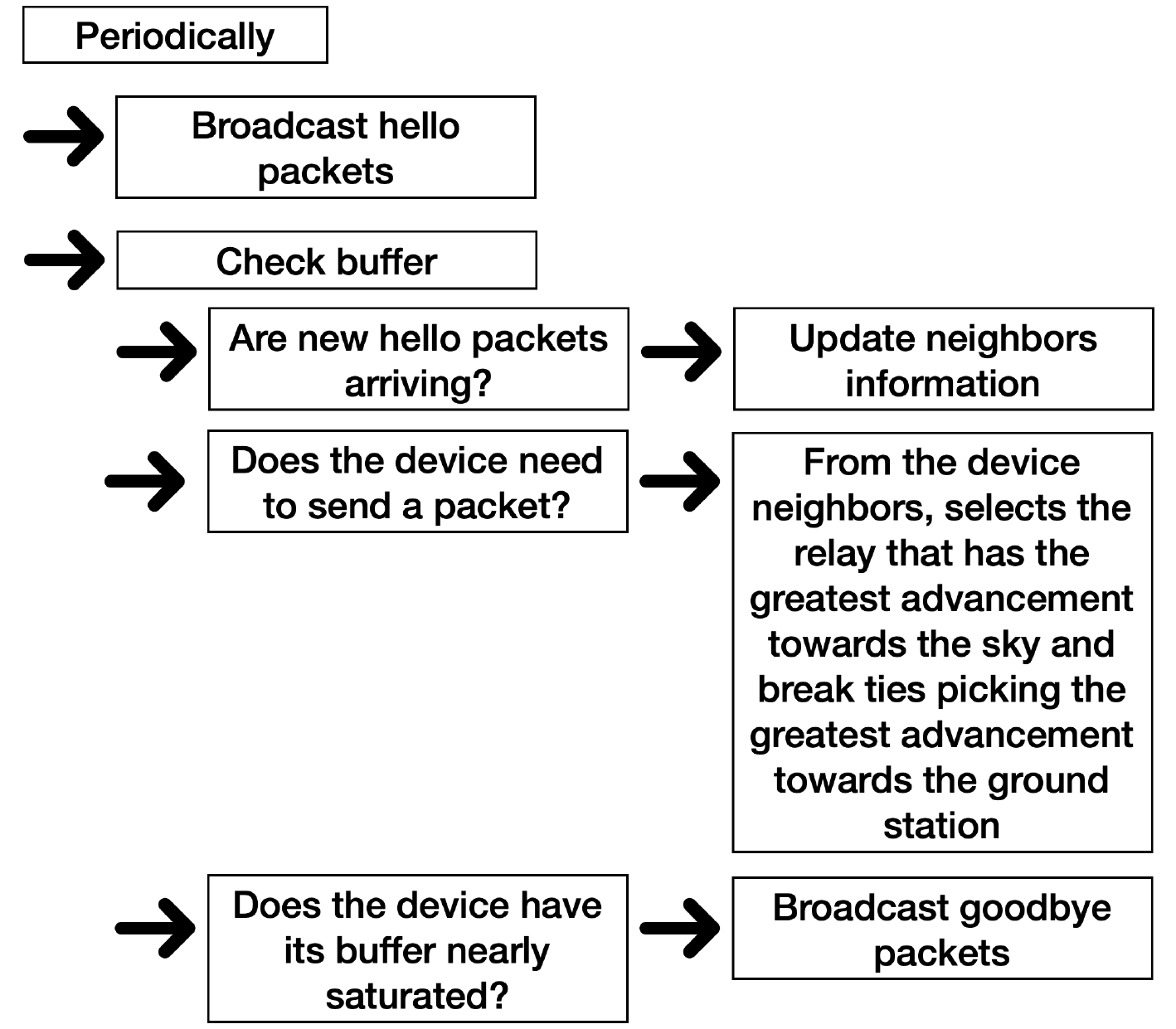}
    \caption{A flow Diagram of the Proposed Mechanism}
    \label{fig:FD}
\end{figure}

A complete flow diagram of the proposed mechanism is presented in Fig. \ref{fig:FD}
\vspace{-0.2cm}

\subsection{Simulation}
We randomly deploy all the nodes in a predefined 3D region (1km$\times$ 1km $\times$ 4km). We keep the depth of the water to 1km, the maximum height of the drones from the water's surface to 1km, and the height of the fixed wings nodes to 3km from the water's surface. We set the transmission range of the underwater nodes to 500m, surface and drone nodes to 1km, and 10 km for the fixed wing nodes, the data rate for the acoustic channel is 10 kbps and the RF link is 6 Mbps, and accordingly, we calculate the transit power of the nodes based on relevant path-loss models \cite{aman2023security}.  
We chose two main performance metrics:  average end-to-end delay and packet delivery ratio (PDR)  to evaluate the proposed cross-media routing mechanism.  Average end-to-end delay is an important metric to evaluate routing mechanisms. Its importance as a metric may vary depending on the specific application and requirements of the systems. Some applications may tolerate higher delays, while others demand near-instantaneous communication. Thus, the selection of appropriate delay thresholds and trade-offs between delay and other metrics should be determined based on the specific needs of the system and its applications. Next, PDR is another important metric in routing problems because it directly reflects the effectiveness and reliability of the routing protocol. It measures the ratio of successfully delivered packets to the total number of packets transmitted within a network.  It helps in maintaining a good quality of service, efficient resource utilization, network reliability, protocol evaluation, and fault detection and troubleshooting.

\begin{figure}
    \centering
    \includegraphics[width=0.3\textwidth]{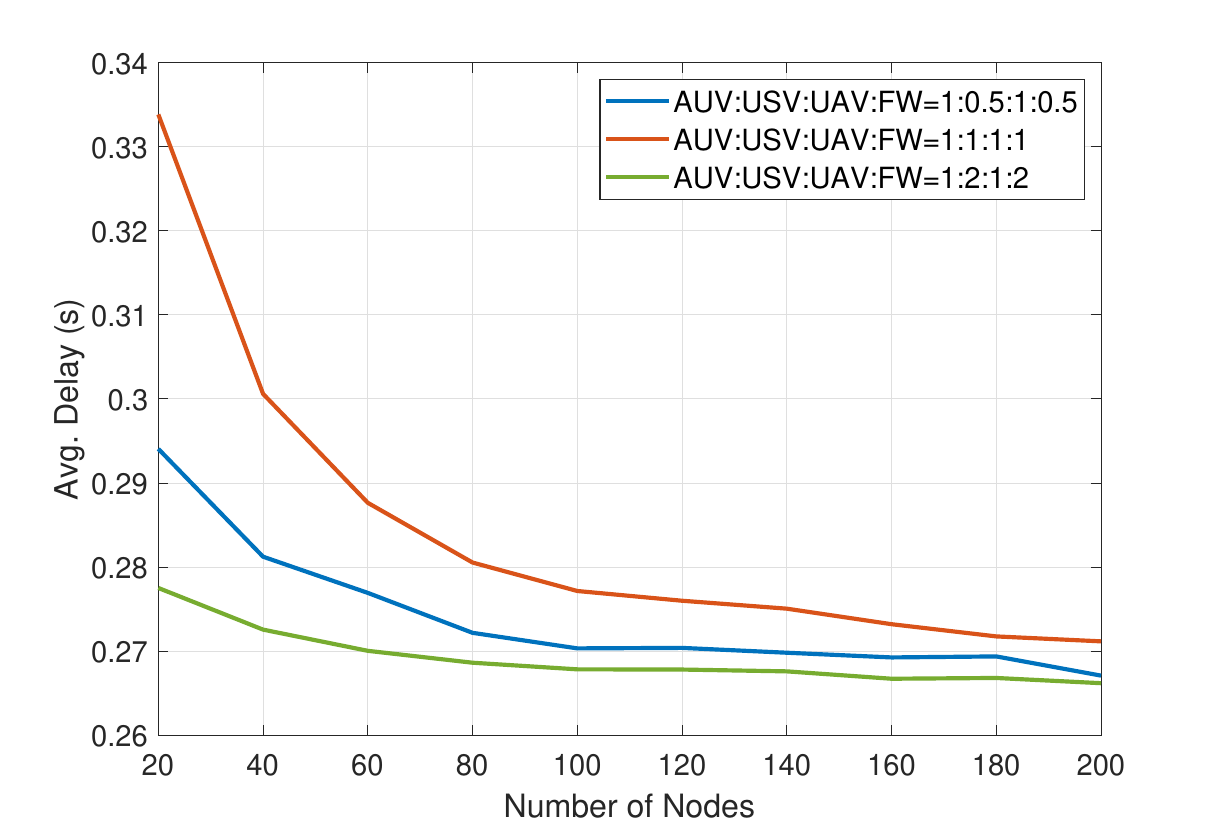}
    \caption{Average Delay against the total number of AUVs. Delay is averaged over 10000 random nodes' deployments.}
    \label{fig:D_v_N}
\end{figure}
Fig \ref{fig:D_v_N} shows the performance of the average end-to-end delay against the number of underwater nodes, note that the total numbers of other nodes are kept according to the ratios mentioned in the legend of the figure. One can clearly see that the end-end delay is not only affected by the number of underwater nodes but the over-water nodes too. This attests to the need for cross-media routing for design purposes that allow the designer to estimate the required configurations for achieving a target delay in the system. Note that three main delay components are considered in this work, i.e., transmission delay (function of hops count, data rate, and packet size), processing delay (the fixed amount of time a node required to process the received packets), and propagation delay (a function of distance and speed of the carrier).

\begin{figure}
    \centering
    \includegraphics[width=0.3\textwidth]{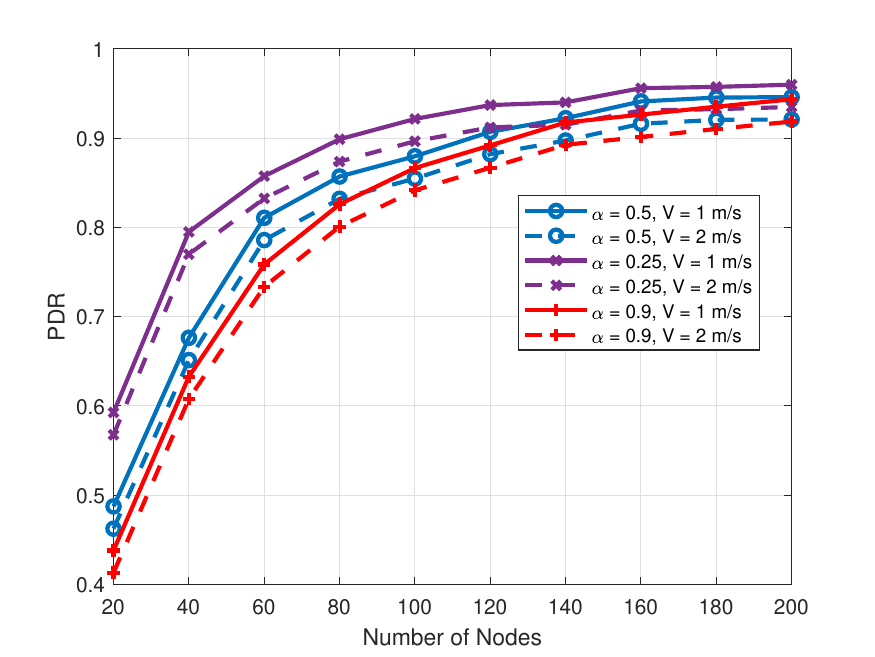}
    \caption{PDR against the total number of AUVs. PDR is averaged over 10000 random nodes' deployments.}
    \label{fig:PDR_v_N}
\end{figure}

Next, Fig. \ref{fig:PDR_v_N} demonstrates the PDR against the number of underwater nodes. We take the 1:05:1:0.5 ratio among the nodes, and for the sake of brevity, the same speed for all the nodes. The important lessons here are the negative impact of speed on PDR and the role of time frame division between the two distinct nature networks. The increase in the speed of nodes increases the void holes and thus more packets dropped, on the other hand, the time allocated ($\alpha T$ where $T$ is the total time frame) to the slow network (underwater network) has comparatively great performance which means more packets are generated and less are dropped. Note that three types of packet dropped scenarios are considered which are packets dropped due to mobility of nodes, buffer overflow, and excessive delay because of CSMA/CA.  However, we believe that more investigations need to be done to address the rate difference challenge at the surface nodes.

\vspace{-0.2cm}

\section{Challanges \& Open Research Issues}\label{sec:ORI}

\subsection{Heterogeneous Communication Characteristics with Dynamic and Unpredictable Environment} Different communication channels in AMSs have diverse characteristics such as varying bandwidth, propagation delays, signal attenuation, and energy constraints. Developing routing algorithms that can efficiently handle these heterogeneous characteristics and optimize data transfer across different media is a significant challenge.
 The marine environment is dynamic and unpredictable, presenting challenges for cross-media routing. Factors such as changing water conditions, underwater obstacles, and signal interference can impact communication performance and the availability of specific media. Routing algorithms need to adapt to these environmental changes and make real-time decisions for effective routing.
\vspace{-0.2cm}

\subsection{Difference of Bandwidths \& Data Rates} Underwater acoustic networks, which are commonly used in AMSs, typically have limited bandwidth and lower data rates compared to satellite or terrestrial networks. Routing algorithms must consider these limitations and ensure efficient utilization of available bandwidth while minimizing latency and maximizing data transfer. The main challenge in cross-media routing is dealing with the inherent differences in communication characteristics and limitations of each medium. For example, underwater acoustic networks have limited bandwidth and higher propagation delays compared to satellite or terrestrial networks. Therefore, routing algorithms need to consider these factors to optimize data transfer and ensure reliable communication across different media. One such challenge has been addressed in the case study but with a heuristic approach. Optimal approaches need to be studied to cope with the bandwidth and rate differences problem. 
\vspace{-0.2cm}

\subsection{Hybrid Underwater Communication Networks}
Such networks also known as multi-modal underwater communication networks where mainly optical and acoustic communication technologies are combined together to get enhanced communication experiences are recently been experimentally developed and tested by the center for maritime research and experimentation (CMRE) \cite{9706085}. The introduction of such systems with multiple acoustic and optical nodes not only provides challenges at the surface nodes but also in the underwater layer. Cross-media routing in the presence of hybrid underwater communication networks needs a thorough investigation. 
\vspace{-0.2cm} 

\subsection{Energy Constraints} AMSs often operate on limited energy resources, and energy-efficient communication is crucial for prolonged mission durations. Cross-media routing algorithms should consider energy constraints and optimize routing decisions to minimize energy consumption while maintaining reliable communication. A Plethora of work has been done to enhance energy efficiency through routing but in single media networks \cite{khan2020underwater}. Energy-efficient cross-media routing needs to get the same attention. Energy is a paramount resource especially for underwater nodes, the PDR at the Aerial or surface layer directly impacts the energy use at the underwater layer. One needs to find the relation between energy use at the underwater layer and other routing parameters (PDR, delay, throughput, etc.) in cross-media routing mechanisms.  
\vspace{-0.2cm}

\subsection{Security and Privacy} Ensuring secure and private communication in AMSs is critical, as they may transmit sensitive data or receive control commands from remote operators. The routing mechanism is prone to various types of attacks such as (i) eavesdropping, (ii) impersonation, (iii) Sybil attacks, and (iv) by malicious sensor nodes located in the vicinity. For instance, a malicious node can launch a wormhole attack (i.e., a prominent attack in the routing mechanism where it advertises itself as the best relay node to others, eavesdrop on a legitimate packet, or impersonate the identity of one or many legitimate underwater sensors by faking its (their) physical address(es). To this end, cross-media routing algorithms need to incorporate security measures to protect data integrity, confidentiality, and authentication against potential threats or malicious attacks.

\subsubsection*{A Naturally-Built Shield: Physical Layer Security (PLS)}
The potential of PLS has been motivating the wireless communications community towards using it as an additional security layer to reinforce overall network security. 

\begin{itemize}
    \item PLS for Confidentiality: Channel impulse or frequency responses can exhibit an inherent source of randomness and uniqueness, from which the two underwater sensors can generate a secret key based on channel (i)  reciprocity and (ii) spatial decorrelation \cite{secrecyacoustic}. Additionally, information-theoretic security can be established in a keyless manner by maximizing the secrecy capacity, defined as the capacity difference between the benign and malign channels, via several techniques, such as jamming, spatial diversity, and transmission scheduling, to cite a few.

    \item Physical Layer Authentication (PLA): The use of unique channel and hardware characteristics can be helpful in authenticating legitimate network nodes and preventing spoofing/impersonation attacks from illegitimate spoofers \cite{waqas2018}.
\end{itemize}
\vspace{-0.2cm}

\subsection{Link Layer Performance}
Routing protocol has a direct relation with the link layer performance. Typically, a buffer or queue is considered at the link layer where packets are arrived due to sensing of the sensor nodes due to other nodes for the given node acting as a relay node.  The amount of data packets coming to the buffer due to sensing is most of the time a constant rate data. Now, the question is what arrival rate or constant arrival rate is optimal or best suited to a network. The answer to the question depends on several system parameters. In this regard, a key performance metric is an effective capacity which is a framework based on the queuing theory that analyzes wireless communication systems at the link layer for randomly time-varying service process \cite{Amjad:CST:2019}. For single media systems, similar works have been witnessed in the past\cite{Wang:ICC:2011} but this problem has not been considered yet for cross-media systems.

\vspace{-0.2cm}

\subsection{Mobility Aware Cross-Media Routing}
 Mobility-aware routing protocols are important to avoid/reduce the damage produced by the mobility of nodes to the performance of the routing mechanism, as can be seen in Fig. \ref{fig:PDR_v_N}. Typically, in mobility-aware protocols, routes are dynamically established and neighbor information is frequently exchanged to avoid the void-hole scenarios. In the context of AMS, such routing protocols have the utmost importance due to the high mobility nature of airborne nodes. Additionally, it's important to study the impact of aerial nodes' mobility on the performance of the underwater part.   

\vspace{-0.2cm}

\subsection{Resource Optimization}
Utilization of the resources of the network plays an important role in meeting many objectives, like a target data rate, delay, PDR, network lifetime, etc. Resource optimization in cross-media routing is essential for achieving efficient, reliable, and sustainable communication across underwater, surface, and aerial networks. By optimizing bandwidth, energy consumption, computational resources, and overall network utilization, cross-media routing systems can maximize performance, adaptability, and resilience, leading to enhanced exploration capabilities, scientific discoveries, and practical applications in various domains such as oceanography, environmental monitoring, and disaster response.

\subsection{Scalability and Network Size} AMSs can involve a large number of nodes, sensors, and vehicles distributed across a wide area. Scalability becomes a challenge when designing cross-media routing algorithms to handle large-scale networks efficiently. Routing protocols should scale well with network size while maintaining low overhead and optimal routing paths.
\vspace{-0.2cm}


\section{Conclusion}\label{sec:conclusion}
In this magazine paper, we have embarked on a comprehensive exploration of cross-media routing, a novel concept that combines marine and aerial interfaces. The journey toward fully realizing the potential of cross-media routing is not without obstacles. We have discussed the challenges that arise in this context, presenting an array of exciting opportunities for future research and innovation. By overcoming these hurdles, we can unlock unprecedented possibilities for cross-media communication and exploration.
As wireless communication continues to expand its reach into underwater and aerial domains, it is imperative to comprehend and address the complexities associated with cross-media routing. The findings presented in this paper serve as a stepping stone towards a future where seamless communication across diverse environments becomes a reality. With further research and innovation, we can chart new frontiers and unleash the full potential of cross-media communication, opening doors to unconventional discoveries and transformative applications.
\vspace{-0.2cm}
\section*{Acknowledgement}
\begin{small}
This work is partially funded by the G5828 ``SeaSec: DroNets for Maritime Border and Port Security" project under the NATO's Science for Peace and Security Programme.
\end{small}
\vspace{-0.2cm}

\bibliographystyle{IEEEtran}
\small
\bibliography{references}
\vspace{-0.2cm}

\section*{Biographies}

\textbf{Waqas Aman} [M] is a postdoctoral researcher at the College of Science and Engineering, Hamad Bin Khalifa University, Doha, Qatar, since Oct. 2022. He completed his Ph.D. in 2021 from Information Technology University on a prestigious scholarship in Pakistan. During his Ph.D., he has been a visiting postgraduate researcher at the University of Glasgow, UK, and a research intern at King Abdullah University of Science and Technology, Saudi Arabia. He also has distinctions in his MS and BS studies and was awarded by \textit{Trans. on Emerg. Tel. Tech.} for his paper to be the top downloaded paper in 2018. He has/is served/serving as a reviewer for IEEE TWC, WCL, Sensors Journal, IoT journal and to name a few, and a session chair for IEEE VTC-Spring-2023.

\textbf{Flavio Giorgi} received his M.Sc. degree in Computer Science in 2022 from Sapienza University of Rome, Italy. He currently is a Ph.D. candidate at Sapienza University. His current research focuses on UAV networks. He consistently  serves as reviewer for several flagship conferences in networking and mobile computing.

\textbf{Giulio Attenni} is currently pursuing a Ph.D. degree in  Computer Science at Sapienza University of Rome, Italy, where he also received his Master's (2022) and Bachelor's (2019) degrees in Computer Science.
His research interests include sustainable mobility, green networking, and algorithmic fairness.
His current research activities focus on route planning and task assignment for networks of aerial drones.

\textbf{Saif Al-Kuwari} [SM] received a Bachelor of Engineering in Computers and Networks from the University of Essex (UK) in 2006 and two PhDs from the University of Bath and Royal Holloway, University of London (UK) in Computer Science, both in 2012. He is currently an assistant professor at the College of Science and Engineering at Hamad Bin Khalifa University. His research interests include applied cryptography, Quantum Computing, Computational Forensics, and their connections with Machine Learning. He is IET and BCS fellow, and IEEE and ACM senior member.

\textbf{Elmehdi Illi} [M]  received the Ph.D. degree from the ENSIAS College of Engineering, Mohammed V University of Rabat in 2019.
From February to May 2022, he was a Visiting Researcher with the College of Science of Engineering, Hamad Bin Khalifa University, Doha, Qatar, where he serves currently as a Postdoctoral Fellow. His main research interests include performance analysis and the design of wireless communication systems.
Dr. Illi was a recipient of the Best Paper Award from the ACOSIS16 Conference. He has been the Registration, Website, and Media Chair for CommNet Conference since 2020. He served as an Organizing Committee Member for various international conferences, such as ACOSIS-16 and CommNet for its last five editions.

\textbf{Marwa Qaraqe} [SM] is an Associate Professor in the Division of Information and Communication Technology in the College of Science and Engineering at Hamad Bin Khalifa University. She received her bachelor`s degree in Electrical Engineering from Texas AM University in Qatar in 2010 and went on to earn her MSc and PhD in Electrical Engineering from Texas AM University in College Station, TX, USA in August of 2012 and May of 2016, respectively. Dr. Qaraqe's research interests cover the following areas: wireless communication, signal processing, and machine learning, and their application in multidisciplinary fields.  Dr. Qaraqe has received several awards throughout her academic career, including the Best Paper Award in the 2022 IEEE Global Communications Conference (GLOBECOM), Richard E. Wing Award for Excellence in Research, and first place in the Qatar Foundation Annual Research Forum in the Health and Biomedical Sector.

\textbf{Gaia Maselli } is an Associate Professor at the Department of Computer Science at Sapienza University of Rome, Italy. She holds a Ph.D. in computer science from the University of Pisa, Italy. Prof. Maselli's current research interests concern design and implementation aspects of mobile networks and wireless communications systems, with particular focus on drones, back-scattering networks and Internet of Things. Other interests include design and performance evaluation of networking protocols for RFID systems. 
She serves as member of the TPC of several international conferences, is associate editor of Elsevier Computer Communications journal, and serves as reviewer for several journals, such as IEEE TMC, IEEE TON. 
She participated to many European Community research projects.

\textbf{Roberto Di Pietro} [F] received the Ph.D. degree in computer science from the University of Roma ``La Sapienza," Rome, Italy, in 2004. He is a Full Professor of Cybersecurity with Hamad Bin Khalifa University-CSE, Doha, Qatar. He was previously with the capacity of Global Head Security Research at Nokia Bell Labs, Murray Hill, NJ, USA, and an Associate Professor of Computer Science with the University of Padova, Padova, Italy.
He has been working in the security field for 25 years, leading both technology-oriented and research-focused teams in the private sector, government, and academia. Prof. Pietro was awarded the Chair of Excellence from the University Carlos III, Madrid, Spain, in 2011 and 2012. He received the Jean-Claude Laprie Award for having significantly influenced the theory and practice of Dependable Computing in 2020, and the Individual Inventor Award by HBKU for outstanding contributions to intellectual property creation and technological advancements in 2022.








\end{document}